\documentclass{llncs}

\usepackage{latexsym}
\usepackage{amssymb}
\usepackage{amsmath}
\usepackage{psfig}
\usepackage{url}

\newcommand{\inb}{\mathit{in\_block}}

\newcommand{\at}{\mathit{At}}

\newcommand{\vd}[1]{\text{vdW}_{#1}}

\begin{document}

\pagestyle{empty}

\title{Satisfiability and computing van der Waerden numbers}

\titlerunning{SAT and van der Waerden numbers}

\author{Michael R. Dransfield\inst{1} \and Victor W. Marek\inst{2} \and
Miros\l aw Truszczy\'nski\inst{2}}
\authorrunning{Michael Dransfield et al.}
\institute{National Security Agency, Information Assurance Directorate,
Ft. Meade, MD 20755\\
\and
Department of Computer Science, University
of Kentucky, Lexington,\\
KY 40506-0046, USA}


\maketitle

\begin{abstract}
In this paper we bring together the areas of combinatorics and
propositional satisfiability. Many combinatorial theorems establish,
often constructively, the existence of positive integer functions,
without actually providing their closed algebraic form or tight lower
and upper bounds. The area of Ramsey theory is especially rich in such 
results. Using the problem of computing van der Waerden numbers as an
example, we show that these problems can be represented by parameterized 
propositional theories in such a way that decisions concerning their
satisfiability determine the numbers (function) in question. We show
that by using general-purpose complete and local-search techniques for testing 
propositional satisfiability, this approach becomes effective ---
competitive with specialized approaches. By following it, we were able 
to obtain several new results pertaining to the problem of computing 
van der Waerden numbers. We also note that due to their properties,
especially their structural simplicity and computational hardness, 
propositional theories that arise in this research can be of use in 
development, testing and benchmarking of SAT solvers.
\end{abstract}

\section{Introduction}\label{intro}

In this paper we discuss how the areas of propositional satisfiability
and combinatorics can help advance each other. On one hand, we show 
that recent dramatic improvements in the efficiency of SAT solvers 
and their extensions make it possible to obtain new results in 
combinatorics simply by encoding problems as propositional theories, 
and then computing their models (or deciding that none exist) using 
off-the-shelf general-purpose SAT solvers. On the other hand, we argue 
that combinatorics is a rich source of structured, parameterized 
families of hard propositional theories, and can provide useful sets 
of benchmarks for developing and testing new generations of SAT 
solvers.

In our paper we focus on the problem of computing van der Waerden 
numbers. The celebrated van der Waerden theorem \cite{vdw27} asserts 
that for every positive integers $k$ and $l$ there is a positive 
integer $m$ such that every partition of $\{1,\ldots,m\}$ into $k$ 
blocks (parts) has at least one block with an arithmetic progression 
of length $l$. The problem is to find the least such number $m$. This
number is called the {\em van der Waerden number} $W(k,l)$. Exact 
values of $W(k,l)$ are known only for five pairs $(k,l)$. For other
combinations of $k$ and $l$ there are some general lower and upper
bounds but they are very coarse and do not give any good idea about 
the actual value of $W(k,l)$. In the paper we show that SAT solvers 
such as POSIT \cite{posit}, and SATO \cite{sato}, 
as well as recently 
developed local-search solver {\em walkaspps} \cite{lt03}, designed 
to compute models for propositional theories  extended by cardinality 
atoms \cite{et01a}, can improve lower bounds for van der Waerden numbers 
for several combinations of parameters $k$ and $l$.

Theories that arise in these investigations are determined by the two
parameters $k$ and $l$. Therefore, they show a substantial degree of
structure and similarity. Moreover, as $k$ and $l$ grow, these theories
quickly become very hard. This hardness is only to some degree an
effect of the growing size of the theories. For the most part, it is
the result of the inherent difficulty of the combinatorial problem in
question. All this suggests that theories resulting from hard
combinatorial problems defined in terms of tuples of integers may serve
as benchmark theories in experiments with SAT solvers.

There are other results similar in spirit to the van der Waerden
theorem. The Schur theorem states that for every positive integer 
$k$ there is an integer $m$ such that every partition of $\{1,
\ldots, m\}$ into $k$ blocks contains a block that is not sum-free. 
Similarly, the Ramsey theorem (which gave name to this whole area 
in combinatorics) \cite{ra28} concerns the existence of monochromatic 
cliques in edge-colored graphs, and the Hales-Jewett theorem 
\cite{hj63} concerns the existence of monochromatic lines in colored 
cubes. 
Each of these results gives rise to a particular function defined on 
pairs or triples of integers and determining the values of these 
functions is a major challenge for combinatorialists. In all cases, 
only few exact values are known and lower and upper estimates are very 
far apart. Many of these results were obtained by means of specialized 
search algorithms highly depending on the combinatorial properties of the
problem. Our paper shows that generic SAT solvers are maturing to the
point where they are competitive and sometimes more effective than
existing advanced specialized approaches.

\section{van der Waerden numbers}
\label{vdw-sub}

In the paper we use the following terminology. By $\mathbb{Z}^+$
we denote the set of positive integers and, for $m \in \mathbb{Z}^+$,
$[m]$ is the set $\{1,\ldots,m\}$. A {\em partition} of a set
$X$ is a collection ${\cal A}$ of nonempty and mutually disjoint 
subsets of $X$ such that $\bigcup {\cal A} = X$. Elements of ${\cal A}$
are commonly called {\em blocks}.

Informally, the van der Waerden theorem \cite{vdw27} states that 
if a sufficiently long initial segment of positive integers is 
partitioned into a few blocks, then one of these blocks has to 
contain an arithmetic progression of a desired length. Formally, the
theorem is usually stated as follows.

\begin{theorem}[van der Waerden theorem] \label{vdw27}
For every $k, l \in \mathbb{Z}^+$, there is $m \in \mathbb{Z}^+$ 
such that for every partition $\{A_1,\ldots,A_k\}$ of $[m]$,
there is $i$, $1 \leq i \leq k$, such that block $A_i$ contains 
an arithmetic progression of length at least $l$.
\end{theorem}

We define the {\em van der Waerden number} $W(k,l)$ to be the least 
number $m$ for which the assertion of Theorem \ref{vdw27} holds. 
Theorem \ref{vdw27} states that van der Waerden numbers are well 
defined.

One can show that for every $k$ and $l$, where $l\geq 2$, $W(k,l) > k$.
In particular, it is easy to see that $W(k,2) = k+1$. From now on, we
focus on the non-trivial case when $l\geq 3$.

Little is known about the numbers $W(k,l)$. In particular, no closed
formula has been identified so far and only five exact values are 
known. They are shown in Table \ref{tab1} \cite{bn79,grs90}.

\begin{table} \label{table.vdw}
\begin{center}
\begin{tabular}{ || l  r | r | r | r ||}
\hline
     & $l$  & $3$ & $4$  & $5$ \\ 
$k$  &      &     &      &     \\ 
\hline
$2$  &      & $9$ & $35$ & $178$ \\
$3$  &      & $27$&      &       \\
$4$  &      & $76$&      &       \\
\hline
\end{tabular}
\end{center}
\caption{Known non-trivial values of van der Waerden numbers}
\label{tab1}
\end{table}

Since we know few exact values for van der Waerden numbers, it is
important to establish good estimates. One can show that the Hales-Jewett
theorem entails the van der Waerden theorem, and some upper bounds for the
numbers $W(k,l)$ can be derived from the Shelah's proof of the former
\cite{sh88}. Recently, Gowers \cite{go01} presented stronger upper
bounds, which he derived from his proof of the Szemer\'edi theorem
\cite{sz75} on arithmetic progressions.

In our work, we focus on lower bounds. Several general results are
known. For instance, Erd\"os and Rado \cite{er52} provided a
non-constructive proof for the inequality
\[
W(k,l) > (2(l-1) k^{l-1})^{1/2}.
\]
For some special values of parameters $k$ and $l$, Berlekamp obtained
better bounds by using properties of finite fields \cite{be68}. These
bounds are still rather weak. His strongest result concerns the case
when $k=2$ and $l-1$ is a prime number. Namely, he proved that when
$l-1$ is a prime number,
\[
W(2,l) > (l-1) 2^{l-1} .
\]
In particular, $W(2,6) > 160$ and $W(2,8) > 896$.

Our goal in this paper is to employ propositional satisfiability solvers 
to find lower bounds for several small van der Waerden numbers. The
bounds we find significantly improve on the ones implied by the results
of Erd\"os and Rado, and Berlekamp. 

We proceed as follows. For each triple of positive integers 
$\langle k,l,m\rangle$, we define a propositional CNF theory 
$\text{vdW}_{k,l,m}$ and then show that $\text{vdW}_{k,l,m}$ is 
satisfiable if and only if $W(k,l) > m$. With such encodings, one can 
use SAT solvers (at least in principle) to determine the satisfiability 
of $\vd{k,l,m}$ and, consequently, find $W(k,l)$. Since $W(k,l) > k$, 
without loss of generality we can restrict our attention to $m > k$.
We also show that more concise encodings are possible, leading
ultimately to better bounds, if we use an extension of propositional
logic by {\em cardinality atoms} and apply to them solvers capable of 
handling such atoms directly.

To describe $\vd{k,l,m}$ we will use a standard first-order language, 
without function symbols, but containing a predicate symbol $\inb$ 
and constants $1,\ldots,m$. An intuitive reading of a ground atom
$\inb(i,b)$ is that an integer $i$ is in block $b$. 

We now define the theory $\vd{k,l,m}$ by including in it the following
clauses:
\newcounter{ct1}
\begin{list}{vdW\arabic{ct1}:\ }{\usecounter{ct1}\topsep 0.1in
\parsep 0in\itemsep 0.05in\itemindent 0in\leftmargin 0.65in\labelwidth
0.5in \labelsep 0.1in}
\item $\neg \inb(i,b_1) \lor \neg \inb(i,b_2)$, \ for every $i\in[m]$ and
every $b_1,b_2\in[k]$ such that $b_1<b_2$,
\item $\inb(i,1) \vee \ldots \vee \inb(i,k)$, \ for every $i\in [m]$,
\item $\neg \inb(i,b) \vee \neg \inb(i+d,b) \vee \ldots \vee \neg
\inb(i+(l-1)d, b)$, \ for every $i,d \in [m]$ such that $i+(l-1)d\leq m$,
and for every $b$ such that $1\leq b\leq k$.
\end{list}

As an aside, we note that we could design $\vd{k,l,m}$ strictly as a 
theory in propositional language using propositional atoms of the form
$\mathit{in\_block}_{i,b}$ instead of ground atoms $\inb(i,b)$. However, 
our approach
opens a possibility to specify this theory as finite (and independent of
data) collections of {\em propositional schemata}, that is, open clauses
in the language of first-order logic without function symbols. Given a
set of appropriate constants (to denote integers and blocks) such
theory, after grounding, coincides with $\vd{k,l,m}$. In fact, we have
defined an appropriate syntax that allows us to specify both data and
schemata and implemented a grounding program {\em psgrnd} \cite{et01a}
that generates their equivalent ground (propositional) representation.
This grounder accepts arithmetic expressions as well as simple regular
expressions, and evaluates and eliminates them according to their 
standard interpretation. Such approach significantly simplifies the
task of developing propositional theories that encode problems, as well as
the use of SAT solvers \cite{et01a}.

Propositional interpretations of the theory $\vd{k,l,m}$ can be
identified with subsets of the set of atoms $\{\inb(i,b)\colon i
\in [m],\ b\in [k]\}$. Namely, a set $M \subseteq \{\inb(i,b)\colon
i\in [m],\ b\in[k]\}$ determines an interpretation in which all atoms
in $M$ are true and all other atoms are false. In the paper we always
assume that interpretations are represented as sets.

It is easy to see that clauses (vdW1) ensure that if $M$ is a model of 
$\vd{k,l,m}$ (that is, is an interpretation satisfying all clauses of 
$\vd{k,l,m}$), then for every $i\in [m]$, $M$ contains at most one atom 
of the form $\inb(i,b)$. Clauses (vdW2) ensure that for every $i\in 
[m]$ there is at least one $b\in [k]$ such that $\inb(i,b)\in M$. In 
other words, clauses (vdW1) and (vdW2) together ensure that if $M$ is 
a model of $\vd{k,l,m}$, then $M$ determines a partition of $[m]$ into 
$k$ blocks. 

The last group of constraints, clauses (vdW3), guarantee that elements
from $[m]$ forming an arithmetic progression of length $l$ do not all
belong to the same block. All these observations imply the following 
result.

\begin{proposition}\label{vdw3}
There is a one-to-one correspondence between models of the formula
$\vd{k,l,m}$ 
and partitions of $[m]$ into $k$ blocks so that no block contains
an arithmetic progression of length $l$. Specifically, an
interpretation $M$ is a model of $\vd{k,l,m}$ if and only if
$\{\{i\in [m] \colon \inb(i,b) \in M\}\colon b\in [k]\}$ is a 
partition of $[m]$ into $k$ blocks such that no block contains an 
arithmetic progression of length $l$.
\end{proposition}

Proposition \ref{vdw3} has the following direct corollary.

\begin{corollary}\label{c.vdw}
For every positive integers $k, l$, and $m$, with $l \geq 2$ and $m > k$,
$m < W(k,l)$ if and only if the formula $\vd{k,l,m}$ is
satisfiable.
\end{corollary}

It is evident that if $m$ has the property that $\vd{k,l,m}$ is
unsatisfiable then for every $m' > m$, $\vd{k,l,m'}$ is also 
unsatisfiable. Thus, Corollary \ref{c.vdw} suggests the following 
algorithm that, given $k$ and $l$, computes the van der Waerden number 
$W(k,l)$: for consecutive integers $m=k+1, k+2,\ldots$ we test whether 
the theory $\vd{k,l,m}$ is satisfiable. If so, we continue. If not, we 
return $m$ and terminate the algorithm. By the van der Waerden theorem, 
this algorithm terminates. 

It is also clear that there are simple symmetries involved in
the van der Waerden problem. If a set $M$ of atoms of the form 
$\inb(i,b)$ is a model of the theory $\vd{k,l,m}$, and $\pi$ is 
a permutation of $[k]$, then the corresponding set of atoms 
$\{\inb(i,\pi(b))\colon \inb(i,b)\in M\}$ is also a model of
$\vd{k,l,m}$, and so is the set of atoms $\{\inb(m+1-i,b)\colon 
\inb(i,b)\in M\}$. 

Following the approach outlined above, adding clauses to break these
symmetries, and applying POSIT  \cite{posit} and SATO \cite{sato}
as a SAT solvers we were able to establish that 
$W(4,3) = 76$ and compute a ``library'' of counterexamples (partitions
with no block containing arithmetic progressions of a specified length)
for $m = 75$. We were also able to find several lower bounds on 
van der Waerden numbers for larger values of $k$ and $m$. 

However, a major limitation of our first approach is that the size of 
theories $\vd{k,l,m}$ grows quickly and makes complete SAT 
solvers ineffective. Let us estimate the size of the theory
$\vd{k,l,m}$. The total size of clauses (vdW1) (measured as the 
number of atom occurrences) is $\Theta(mk^2)$. The size of clauses 
(vdW2) is $\Theta(mk)$. Finally, the size of clauses (vdW3) is 
$\Theta(m^2)$ (indeed, there are $\Theta(m^2/l)$ arithmetic 
progressions of length $l$ in $[m])$\footnote{Goldstein \cite{dg02}
provided a 
precise formula. When $r = rm(m-1,l-1)$ and $q = q(m-1,l-1)$ then 
there are $q\cdot r + \binom{q-1}{2}\cdot (l-1)$ arithmetic 
progressions of length $l$ in $[m]$.}. Thus, the total size of the 
theory $\vd{k,l,m}$ is $\Theta(mk^2+m^2)$.

To overcome this obstacle, we used a two-pronged 
approach. First, as a modeling language we used PS+ logic \cite{et01a},
which is an extension of propositional logic by cardinality atoms. 
Cardinality atoms support concise representations of constraints of the
form ``at least $p$ and at most $r$ elements in a set are true'' and
result in theories of smaller size.
Second, we used a local-search algorithm, {\em walkaspps}, for finding 
models of theories in logic PS+ that we have designed and implemented
recently \cite{lt03}. Using encodings as theories in logic PS+ and {\em
walkaspps} as a solver, we were able to obtain substantially stronger 
lower bounds for van der Waerden numbers than those know to date.

We will now describe this alternative approach. 
For a detailed treatment of the PS+
logic we refer the reader to \cite{et01a}. In this paper, we will
only review most basic ideas underlying the logic PS+ (in its
propositional form). By a {\em propositional cardinality atom} ({\em
c-atom} for short), we mean any expression of the form $m\{p_1,\ldots,
p_k\}n$ (one of $m$ and $n$, but not both, may be missing), where $m$ 
and $n$ are non-negative integers and $p_1 ,\ldots,p_k$ are 
propositional atoms from $\at$. The notion of a clause generalizes in 
an obvious way to the language with cardinality atoms. Namely, a {\em 
c-clause} is an expression of the form
\begin{equation}\label{eq.cl}
C=\ \ A_1\vee \ldots \vee A_s\vee \neg B_1\vee\ldots\vee \neg B_t,
\end{equation}
where all $A_i$ and $B_i$ are (propositional) atoms or cardinality
atoms.

Let $M\subseteq \at$ be a set of atoms. We say that $M$ {\em satisfies}
a cardinality atom $m\{p_1,\ldots,p_k\}n$ if
\[
m \leq |M\cap \{p_1,\ldots,p_k\}|\leq n.
\]
If $m$ is missing, we only require that $|M\cap \{p_1,\ldots,p_k\}|\leq
n$. Similarly, when $n$ is missing, we only require that $m \leq |M\cap
\{p_1,\ldots,p_k\}|$. A set of atoms $M$ {\em satisfies} a c-clause $C$
of the form (\ref{eq.cl})
if $M$ satisfies at least one atom $A_i$ or does not satisfy at least
one atom $B_j$. W note that the expression $1\{p_1,\ldots ,p_k\}1$
expresses the quantifier ``There exists exactly one ...'' - commonly
used in mathematical statements.

It is now clear that all clauses (vdW1) and (vdW2) from $\vd{k,l,m}$ 
can be represented in a more concise way by the following
collection of c-clauses:
\newcounter{ct2}
\begin{list}{vdW$'$\arabic{ct2}:\ }{\usecounter{ct2}\topsep 0.03in
\parsep 0in\itemsep 0in}
\item $1\{\inb(i,1),\ldots,\inb(i,k)\}1$, for every $i\in [m]$.
\end{list}

Indeed, c-clauses (vdW$'$1) enforce that their models, for every
$i\in[m]$ contain exactly one atom of the form $\inb(i,b)$ ---
precisely the same effect as that of clauses (vdW1) and (vdW2).
Let $\vd{k,l,m}'$ be a PS+ theory consisting of clauses (vdW$'$1)
and (vdW3). It follows that Proposition \ref{vdw3} and Corollary 
\ref{c.vdw} can be reformulated by replacing $\vd{k,l,m}$ with
$\vd{k,l,m}'$ in their statements. Consequently, any algorithm
for finding models of PS+ theories can be used to compute van der
Waerden numbers (or, at least, some bounds for them) in the way we
described above.

The adoption of cardinality atoms leads to a more concise 
representation of the problem. While, as we discussed above,  
the size of all clauses (vdW1) and (vdW2) is $\Theta (mk^2 + mk)$, 
the size of clauses (vdW$'$1) is $\Theta(mk)$.

In our experiments, for various lower bound results,
we used the local-search algorithm {\em walkaspps}
\cite{lt03}. This algorithm is based on the same ideas as {\em
walksat} \cite{skc94}. A major difference is that due to the presence of
c-atoms in c-clauses {\em walkaspps} uses different formulas to 
calculate the breakcount and proposes several other heuristics designed
specifically to handle c-atoms.

\section{Results} \label{results}

Our goal is to establish lower bounds for small van der Waerden numbers
by exploiting propositional satisfiability solvers.  Here is a summary
of our results.

\begin{enumerate}
\item Using complete SAT solvers POSIT and SATO and the encoding of the
problem as $\vd{k,l,m}$, we found a ``library'' of all (up to obvious
symmetries) counterexamples to the fact that $W(4,3) > 75$. There are
30 of them. We list two of them in the appendix. A complete list can be
found at \url{http://www.cs.uky.edu/ai/vdw/}. Since there 
are $48$ symmetries, 
of the types discussed above,
the full library of counterexamples consists of 
$1440$  partitions.
\item We found that the formula $\vd{4,3,76}$ is unsatisfiable. Hence,
we found that a ``generic'' SAT solver is capable of finding that 
$W(4,3) = 76$.
\item We established several new lower bounds for the numbers $W(k,l)$. 
They are presented in Table \ref{table.vdw1}. Partitions demonstrating 
that $W(2,8) > 1295$, $W(3,5) > 650$, and $W(4,4) > 408$ are included 
in the appendix. Counterexample partitions for all other inequalities 
are available at
\url{http://www.cs.uky.edu/ai/vdw/}.
We note that our bounds for $W(2,6)$ and $W(2,8)$ are much stronger
than those implied by the results of Berlekamp \cite{be68}, which we 
stated earlier.
\end{enumerate}

\begin{table} \label{table.vdw1}
\caption{Extended results on van der Waerden numbers}
\begin{center}
\begin{tabular}{ || l  r | r | r | r | r | r | r ||}
\hline
     & $l$  & $3$    & $4$    & $5$    & $6$    & $7$    & $8$     \\
$k$  &      &        &        &        &        &        &         \\
\hline
$2$  &      & $9$    & $35$   & $178$  & $>341$ & $>604$ & $>1295$ \\
$3$  &      & $27$   & $>193$ & $>650$ &        &        &         \\
$4$  &      & $76$   & $>408$ &        &        &        &         \\
$5$  &      & $>125$ &        &        &        &        &         \\
$6$  &      & $>180$ &        &        &        &        &         \\
\hline
\end{tabular}
\end{center}
\end{table}

To provide some insight into the complexity of the satisfiability 
problems involved, in Table \ref{table.vdw2} we list the number of
atoms and the number of clauses in the theories $\vd{k,l,m}'$.
Specifically, the entry $k,l$ in this table contains the number of
atoms and the number of clauses in the theories $\vd{k,l,m}'$, where
$m$ is the value given in the entry $k,l$ in Table \ref{table.vdw1}. 

\begin{table} \label{table.vdw2}
\caption{Numbers of atoms and clauses in theories $\vd{k,l,m}'$, used
to establish the results presented in Table \ref{table.vdw1}.}
\begin{center}
\begin{tabular}{ || l  r | r | r | r | r | r | r ||}
\hline
    & $l$  & $3$    & $4$    & $5$    & $6$    & $7$    & $8$    \\
$k$ &   &        &        &        &        &        &        \\
\hline
$2$ &   & 18, 41  & 70, 409 & 356, 7922 & 682, 23257 & 1208, 60804 & 2590, 239575\\
$3$ &   & 108, 534 & 579, 18529 & 1950, 158114 &        &        &        \\
$4$ &   & 304, 5700 & 1632, 110568 &        &        &        &        \\
$5$ &   & 625, 19345 &        &        &        &        &        \\
$6$ &   & 1080, 48240 &        &        &        &        &        \\
\hline
\end{tabular}
\end{center}
\end{table}

\section{Discussion}\label{conclude}

Recent progress in the development of SAT solvers provides an important
tool for researchers looking for both the existence and non-existence
of various combinatorial objects. We have demonstrated that several 
classical questions related to van der Waerden numbers can be naturally 
cast as questions on the existence of satisfying valuations for some 
propositional CNF-formulas. 

Computing combinatorial objects such as van der Waerden numbers is
hard. They are structured but as we pointed out few values are known, 
and new results are hard to obtain. Thus, the computation of those 
numbers can serve as a benchmark (`can we find the configuration 
such that...') for complete and local-search methods, and as a 
challenge (`can we show that a configuration such that ...' does not 
exist) for complete SAT solvers. Moreover, with powerful SAT solvers 
it is likely that the bounds obtained by computation of counterexamples 
are ``sharp'' in the sense that when a configuration is not found then 
none exist. For instance it is likely that $W(5,3)$ is close to 126 
(possibly, it is 126), because 125 was the last integer where we were 
able to find a counterexample despite significant computational effort. 
This claim is further supported by the fact that in all examples where 
exact values are known, our local-search algorithm was able to find 
counterexample partitions for the last possible value of $m$. The 
lower-bounds results of this sort may constitute an important clue for 
researchers looking for nonexistence arguments and, ultimately, for the
closed form of van der Waerden numbers.

A major impetus for the recent progress of SAT solvers comes from 
applications in computer engineering. In fact, several leading SAT 
solvers such as zCHAFF \cite{mmzzm01} and {\em berkmin} \cite{berkmin}
have been developed with the express goal of aiding engineers in 
correctly designing and implementing digital circuits. Yet, the fact 
that these solvers are able to deal with hard optimization problems 
in one area (hardware design and verification) carries the promise 
that they will be of use in another area --- combinatorial 
optimization. Our results indicate that it is likely to be the 
case.

The current capabilities of SAT solvers has allowed us to handle
large instances of these problems. Better heuristics and other
techniques for pruning the search space will undoubtedly further expand
the scope of applicability of generic SAT solvers to problems that,
until recently, could only be solved using specialized software.

\section*{Acknowledgments}
The authors thank Lengning Liu for developing software facilitating
our experimental work. This research has been supported by the Center 
for Communication Research, La Jolla. During the research reported in 
this paper the second and third authors have been partially supported 
by an NSF grant  IIS-0097278.

\section*{Appendix}

Using a complete SAT solver we computed
the library of all partitions (up to isomorphism) of $[75]$ showing 
that $75 < W(4,3)$. Two of these 30 partitions are shown below:\\
\\
\noindent
Solution 1:\\
{\small
Block 1: 6 7 9 14 18 20 23 24 36 38 43 44 46 51 55 57 60 61 73 75\\
Block 2: 4 5 12 22 26 28 29 31 37 41 42 49 59 63 65 66 68 74 \\
Block 3: 1 2 8 10 11 13 17 27 34 35 39 45 47 48 50 54 64 71 72 \\    
Block 4: 3 15 16 19 21 25 30 32 33 40 52 53 56 58 62 67 69 70} \\
\ \\ 
Solution 2:\\ 
{\small
Block 1: 6 7 9 14 18 20 23 24 36 38 43 44 46 51 55 57 60 61 73 \\
Block 2: 4 5 12 22 26 28 29 31 37 41 42 49 59 63 65 66 68 74\\ 
Block 3: 1 2 8 10 11 13 17 27 34 35 39 45 47 48 50 54 64 71 72\\
Block 4: 3 15 16 19 21 25 30 32 33 40 52 53 56 58 62 67 69 70 75}\\ 
\ \\
These two and the remaining 28 partitions can be found at
\url{http://www.cs.uky.edu/ai/vdw/} \\
\ \\ 
\noindent
Next, we exhibit a partition of $[1295]$ into two blocks demonstrating
that $W(2,8)$ \\
$> 1295$.\\
\ \\
{\small
\noindent
Block 1:\\
1 3 4 5 7 8 10 11 13 14 15 16 17 18 21 26 27 29 31 35 38 40 42 43 45 46 51 53 56 62 63 64 67 68 69 71 73 74 75 77 79 80 83 85 86 88 90 94 96 97 98 101 102 103 104 107 110 112 114 116 118 120 123 124 125 130 131 132 135 138 139 142 145 149 152 153 155 157 159 160 161 163 165 166 169 170 171 174 178 179 181 187 188 189 190 192 193 195 198 200 202 205 207 208 209 210 211 212 213 215 216 221 222 224 225 226 228 229 231 232 236 241 247 249 252 253 254 255 259 260 261 262 264 267 268 269 270 272 274 277 278 279 286 288 290 292 293 294 295 296 297 298 301 306 308 309 311 312 313 317 319 320 321 322 323 326 327 328 334 335 336 338 342 346 349 356 358 359 360 367 368 369 370 373 374 377 378 379 382 383 384 385 386 388 395 396 398 399 400 401 402 404 405 408 410 413 414 416 417 420 423 424 426 429 430 433 434 436 437 443 445 446 447 448 449 451 452 453 456 459 463 464 467 469 470 473 475 476 477 478 479 481 485 486 487 488 490 491 494 495 497 499 502 503 504 505 507 508 510 513 515 518 521 522 528 529 530 533 534 539 540 542 546 547 550 555 558 559 560 561 564 571 577 578 579 580 581 583 584 587 589 590 591 594 595 596 597 601 609 611 612 613 614 615 616 618 619 623 624 625 626 627 628 632 634 636 637 639 640 642 643 647 648 651 652 653 660 661 662 663 665 666 668 670 674 675 676 677 678 680 681 683 684 687 688 690 694 695 696 697 698 700 701 702 703 704 706 709 710 715 717 718 722 725 726 727 728 734 739 742 743 744 746 748 752 753 755 756 757 759 763 766 768 770 771 774 775 776 779 781 788 792 795 796 799 801 802 806 807 809 812 816 817 818 819 821 825 826 832 833 835 836 840 841 843 844 845 846 847 848 852 853 855 856 859 862 863 864 867 868 871 872 874 875 876 877 879 881 882 885 886 893 897 898 899 901 902 903 904 905 906 908 909 910 913 915 917 922 923 925 927 928 929 930 931 932 936 937 939 940 941 944 946 947 948 951 952 954 957 960 961 963 964 965 966 967 974 977 982 983 984 986 989 990 993 994 1001 1003 1004 1008 1009 1010 1012 1013 1016 1017 1020 1022 1023 1025 1026 1028 1029 1033 1034 1036 1037 1038 1040 1045 1047 1050 1051 1052 1053 1058 1060 1065 1070 1073 1074 1075 1076 1077 1079 1083 1085 1087 1088 1089 1090 1091 1092 1094 1095 1096 1097 1098 1102 1103 1105 1106 1109 1111 1113 1116 1117 1118 1119 1121 1123 1124 1126 1129 1130 1133 1135 1139 1140 1141 1144 1150 1151 1152 1154 1155 1156 1157 1159 1161 1168 1170 1171 1174 1175 1179 1180 1184 1185 1186 1188 1189 1190 1191 1194 1196 1197 1200 1202 1205 1206 1213 1216 1217 1218 1219 1220 1221 1222 1224 1226 1227 1229 1234 1236 1237 1238 1239 1246 1247 1249 1251 1253 1257 1260 1261 1262 1263 1264 1268 1269 1272 1274 1275 1276 1278 1279 1283 1285 1286 1287 1288 1289 1290 1291 1294 1295\\
Block 2:\\
2 6 9 12 19 20 22 23 24 25 28 30 32 33 34 36 37 39 41 44 47 48 49 50 52 54 55 57 58 59 60 61 65 66 70 72 76 78 81 82 84 87 89 91 92 93 95 99 100 105 106 108 109 111 113 115 117 119 121 122 126 127 128 129 133 134 136 137 140 141 143 144 146 147 148 150 151 154 156 158 162 164 167 168 172 173 175 176 177 180 182 183 184 185 186 191 194 196 197 199 201 203 204 206 214 217 218 219 220 223 227 230 233 234 235 237 238 239 240 242 243 244 245 246 248 250 251 256 257 258 263 265 266 271 273 275 276 280 281 282 283 284 285 287 289 291 299 300 302 303 304 305 307 310 314 315 316 318 324 325 329 330 331 332 333 337 339 340 341 343 344 345 347 348 350 351 352 353 354 355 357 361 362 363 364 365 366 371 372 375 376 380 381 387 389 390 391 392 393 394 397 403 406 407 409 411 412 415 418 419 421 422 425 427 428 431 432 435 438 439 440 441 442 444 450 454 455 457 458 460 461 462 465 466 468 471 472 474 480 482 483 484 489 492 493 496 498 500 501 506 509 511 512 514 516 517 519 520 523 524 525 526 527 531 532 535 536 537 538 541 543 544 545 548 549 551 552 553 554 556 557 562 563 565 566 567 568 569 570 572 573 574 575 576 582 585 586 588 592 593 598 599 600 602 603 604 605 606 607 608 610 617 620 621 622 629 630 631 633 635 638 641 644 645 646 649 650 654 655 656 657 658 659 664 667 669 671 672 673 679 682 685 686 689 691 692 693 699 705 707 708 711 712 713 714 716 719 720 721 723 724 729 730 731 732 733 735 736 737 738 740 741 745 747 749 750 751 754 758 760 761 762 764 765 767 769 772 773 777 778 780 782 783 784 785 786 787 789 790 791 793 794 797 798 800 803 804 805 808 810 811 813 814 815 820 822 823 824 827 828 829 830 831 834 837 838 839 842 849 850 851 854 857 858 860 861 865 866 869 870 873 878 880 883 884 887 888 889 890 891 892 894 895 896 900 907 911 912 914 916 918 919 920 921 924 926 933 934 935 938 942 943 945 949 950 953 955 956 958 959 962 968 969 970 971 972 973 975 976 978 979 980 981 985 987 988 991 992 995 996 997 998 999 1000 1002 1005 1006 1007 1011 1014 1015 1018 1019 1021 1024 1027 1030 1031 1032 1035 1039 1041 1042 1043 1044 1046 1048 1049 1054 1055 1056 1057 1059 1061 1062 1063 1064 1066 1067 1068 1069 1071 1072 1078 1080 1081 1082 1084 1086 1093 1099 1100 1101 1104 1107 1108 1110 1112 1114 1115 1120 1122 1125 1127 1128 1131 1132 1134 1136 1137 1138 1142 1143 1145 1146 1147 1148 1149 1153 1158 1160 1162 1163 1164 1165 1166 1167 1169 1172 1173 1176 1177 1178 1181 1182 1183 1187 1192 1193 1195 1198 1199 1201 1203 1204 1207 1208 1209 1210 1211 1212 1214 1215 1223 1225 1228 1230 1231 1232 1233 1235 1240 1241 1242 1243 1244 1245 1248 1250 1252 1254 1255 1256 1258 1259 1265 1266 1267 1270 1271 1273 1277 1280 1281 1282 1284 1292 1293
\ \\
}
\ \\
\noindent
Next, we exhibit a partition of $[650]$ into three blocks demonstrating
that $W(3,5)$ \\
$> 650$.\\
\ \\
{\small
\noindent
Block 1:\\
1 2 5 6 10 16 18 21 22 23 27 28 31 35 40 44 45 46 55 56 58 59 67 69 73 75 81 82 84 85 86 95 96 97 100 102 103 105 107 110 111 117 121 122 127 130 131 132 133 136 138 141 142 147 148 152 155 156 157 158 163 165 168 171 175 180 181 183 185 186 189 203 207 210 211 212 215 216 218 221 223 225 227 236 238 240 241 242 247 250 252 254 256 259 260 261 262 266 271 277 280 282 287 288 290 291 292 296 300 302 306 310 328 330 331 334 340 345 346 347 348 350 355 362 365 366 367 371 374 375 378 380 383 384 386 390 392 393 395 396 397 399 400 405 407 408 411 412 413 422 433 435 436 439 443 448 449 453 455 456 457 460 463 472 481 485 486 491 493 500 503 505 506 508 509 511 515 517 521 524 525 528 530 532 535 543 548 550 551 552 560 561 565 566 568 569 571 575 583 585 587 596 597 598 607 608 610 616 620 624 625 626 629 630 640 641 642 646\\
\noindent
Block 2:\\
3 4 7 8 9 12 15 24 26 29 32 34 37 39 42 43 49 51 60 61 63 65 68 70 71 74 76 78 79 80 83 87 89 90 91 94 109 112 113 115 118 120 129 134 135 139 140 143 145 149 153 159 160 162 164 167 172 173 176 177 178 179 188 190 195 197 200 205 209 213 214 217 219 220 222 224 230 232 233 234 235 239 244 245 248 249 253 255 270 273 275 279 281 284 285 286 297 299 301 305 308 315 318 323 324 325 327 332 333 335 336 338 339 342 343 344 349 354 356 357 358 360 361 364 368 369 370 377 379 382 385 387 389 394 398 410 415 418 424 425 426 430 432 437 440 445 446 450 452 458 461 465 468 471 474 475 476 480 482 483 487 488 490 492 495 496 499 504 514 519 520 523 526 527 529 534 537 539 540 545 549 555 558 567 570 572 574 577 579 580 581 582 584 588 590 593 599 600 602 604 605 611 612 614 615 618 619 633 636 637 639 644 645 648\\
\noindent
Block 3:\\
11 13 14 17 19 20 25 30 33 36 38 41 47 48 50 52 53 54 57 62 64 66 72 77 88 92 93 98 99 101 104 106 108 114 116 119 123 124 125 126 128 137 144 146 150 151 154 161 166 169 170 174 182 184 187 191 192 193 194 196 198 199 201 202 204 206 208 226 228 229 231 237 243 246 251 257 258 263 264 265 267 268 269 272 274 276 278 283 289 293 294 295 298 303 304 307 309 311 312 313 314 316 317 319 320 321 322 326 329 337 341 351 352 353 359 363 372 373 376 381 388 391 401 402 403 404 406 409 414 416 417 419 420 421 423 427 428 429 431 434 438 441 442 444 447 451 454 459 462 464 466 467 469 470 473 477 478 479 484 489 494 497 498 501 502 507 510 512 513 516 518 522 531 533 536 538 541 542 544 546 547 553 554 556 557 559 562 563 564 573 576 578 586 589 591 592 594 595 601 603 606 609 613 617 621 622 623 627 628 631 632 634 635 638 643 647 649 650
\ \\
}
\ \\
\noindent
Finally, we exhibit a partition of $[408]$ into four blocks demonstrating
that $W(4,4) > 408$.\\
\ \\
{\small
\noindent
Block 1:\\
2 8 11 17 19 20 23 30 38 42 48 50 52 59 61 65 67 71 78 82 83 85 89 90 98 104 107 108 113 119 120 124 127 129 140 143 144 147 150 152 157 158 163 166 181 183 184 198 199 204 214 220 223 226 231 237 240 241 244 250 251 253 259 264 266 270 271 273 278 282 286 287 289 306 312 314 317 318 321 327 329 331 348 351 354 359 361 362 363 366 373 377 378 382 383 386 399 401 402 403 406

\noindent
Block 2:\\
1 3 7 13 15 16 24 26 28 37 39 47 49 57 58 66 73 76 77 81 84 86 87 92 93 94 103 110 111 117 118 121 122 123 125 133 135 151 153 154 155 161 162 167 170 172 176 182 190 194 195 196 207 210 216 228 232 233 234 242 243 245 246 248 249 254 255 256 258 262 275 280 283 284 290 293 297 298 299 305 307 309 328 333 336 341 346 352 353 355 356 358 368 370 371 372 381 385 391 393 404

\noindent
Block 3:\\
4 6 21 22 27 29 31 32 34 35 40 41 44 56 62 63 69 70 72 74 75 79 95 96 99 101 105 109 114 115 116 126 132 134 136 141 145 159 160 165 169 171 174 175 179 180 187 188 191 192 197 200 201 208 209 212 217 219 221 227 229 235 236 247 257 263 267 269 272 274 276 281 291 292 294 300 302 304 310 311 322 324 325 330 332 334 339 340 342 344 345 350 365 367 376 379 388 390 394 397 398 400 407

\noindent
Block 4:\\
5 9 10 12 14 18 25 33 36 43 45 46 51 53 54 55 60 64 68 80 88 91 97 100 102 106 112 128 130 131 137 138 139 142 146 148 149 156 164 168 173 177 178 185 186 189 193 202 203 205 206 211 213 215 218 222 224 225 230 238 239 252 260 261 265 268 277 279 285 288 295 296 301 303 308 313 315 316 319 320 323 326 335 337 338 343 347 349 357 360 364 369 374 375 380 384 387 389 392 395 396 405 408
}\\
\ \\
\noindent
Configurations showing the validity of other lower bounds listed in 
Table \ref{table.vdw1} are available at 
\url{http://www.cs.uky.edu/ai/vdw/}.

\begin{thebibliography}{10}


\bibitem{bn79}
M.D. Beeler and P.E. O'Neil.
\newblock  {Some new van der Waerden numbers}, {\em Discrete
  Mathematics}, 28:135--146, 1979.

\bibitem{be68} E. Berlekamp. A construction for partitions which avoid
long arithmetic progressions. {\em Canadian Mathematical Bulletin} 
11:409--414, 1968.



\bibitem{dp60}
M.~Davis and H.~Putnam.
\newblock  A computing procedure for quantification theory, {\em
  Journal of the Association for Computing Machinery}, 7:201--215, 1960.

\bibitem{et01a}
D. East and M. Truszczy\'nski. Propositional satisfiability in answer-set
programming. Proceedings of Joint German/Austrian Conference on
Artificial Intelligence, KI'2001. Lecture Notes in Artificial
Intelligence, Springer Verlag 2174, pages 138--153. (Full version
available at \url{http://xxx.lanl.gov/ps/cs.LO/0211033}). 2001.

\bibitem{er52}
P.~{Erd\"os} and R.~Rado.
\newblock  Combinatorial theorems on classifications of subsets of
a given set, {\em Proceedings of London Mathematical Society}, 2:417--439,
1952.

\bibitem{posit}
J.W. Freeman.
\newblock  {\em {Improvements to propositional satisfiability
  search algorithms}}, PhD thesis, Department of Computer Science, University
  of Pennsylvania, 1995.

%

\bibitem{berkmin}  
E. Goldberg, Y. Novikov. 
BerkMin: a Fast and Robust SAT-Solver. DATE-2002, pages 142--149, 2002.

\bibitem{dg02}
D. Goldstein.
\newblock Personal communication, 2002.

\bibitem{go01} T. Gowers. A new proof of Szemer\'edi theorem. {\em
Geometric and Functional Analysis}, 11:465-588, 2001.

\bibitem{grs90}
R.L. Graham, B.L. Rotschild, and J.H. Spencer.
\newblock  {\em Ramsey Theory}, Wiley, 1990.

\bibitem{hj63}
A.~Hales and R.I. Jewett.
\newblock  Regularity and positional games, {\em Transactions of
  American Mathematical Society}, 106:222--229, 1963.

\bibitem{jw90}
R.E. Jeroslaw and J.~Wang.
\newblock  solving propositional satisfiability problems, {\em
  Annals of Mathematics and Artificial Intelligence}, 1:167--187, 1990.

\bibitem{lt03}
L. Liu and M. Truszczy\'nski. Local-search techniques in propositional
logic extended with cardinality atoms. In preparation.


%
\bibitem{ss99}
J.P. Marques-Silva and K.A. Sakallah.
\newblock  {GRASP}: {A} new search algorithm for satisfiability,
  {\em {IEEE Transactions on Computers}}, 48:506--521, 1999.


%
\bibitem{mmzzm01}
M.W. Moskewicz, C.F. Magidan, Y.~Zhao, L.~Zhang, and S.~Malik.
\newblock  Chaff: engineering an efficient {SAT} solver, in {\em
  SAT 2001}, 2001.

%
\bibitem{ra28}
F.P. Ramsey.
\newblock  On a problem of formal logic, {\em Proceedings of London
Mathematical Society}, 30:264--286, 1928.

\bibitem{skc94}
B. Selman, H.A. Kautz, and B. Cohen.
\newblock Noise Strategies for Improving Local Search. {\em Proceedings
of AAAI'94}, pp. 337-343. MIT Press 1994.

\bibitem{sh88}
S.~Shelah.
\newblock  {Primitive recursive bounds for van der Waerden
  numbers}, {\em Journal of American Mathematical Society}, 1:683--697, 1988.

\bibitem{sz75}
E. Szemer{\'e}di.
\newblock {On sets of integers containing no $k$ elements in arithmetic
progression}, {\em Acta Arithmetica}, 27:199--243, 1975.

\bibitem{vdw27}
B.L. van~der Waerden.
\newblock  {Beweis einer Baudetschen Vermutung}, {\em Nieuwe
  Archief voor Wiskunde}, 15:212--216, 1927.

\bibitem{sato}
H.~Zhang.
\newblock  {SATO: An efficient propositional prover}, in {\em
  Proceedings of CADE-17}, pages 308--312, 1997.
\newblock Springer Lecture Notes in Artificial Intelligence 1104.

\def\BibNewMarkings{2}
\end{thebibliography}
\end{document}